\documentclass[fleqn,usenatbib]{mnras}
\usepackage{newtxtext,newtxmath}
\usepackage[T1]{fontenc}

\usepackage{amsmath}
\usepackage{graphicx}
\usepackage{xcolor}

\DeclareRobustCommand{\VAN}[3]{#2}
\let\VANthebibliography\thebibliography
\def\thebibliography{\DeclareRobustCommand{\VAN}[3]{##3}\VANthebibliography}

\defcitealias{ocker2022}{O22}

\title[FRB Scattering Variations]{Scattering variability detected from the circumsource medium of FRB~20190520B} 

\author[S.K. Ocker et al.]{Stella Koch Ocker,$^{1}$\thanks{E-mail: sko36@cornell.edu}
James M. Cordes,$^{1}$
Shami Chatterjee,$^{1}$
Di Li,$^{2,3}$
Chen-Hui Niu,$^{2}$
James W. McKee,$^{4,5}$ \newauthor
Casey J. Law,$^{6,7}$
and Reshma Anna-Thomas$^{8}$
\\
$^{1}$Department of Astronomy and Cornell Center for Astrophysics and Planetary Science, Cornell University, Ithaca, NY 14850, USA\\
$^{2}$National Astronomical Observatories, Chinese Academy of Sciences, Beijing 100101, China\\
$^{3}$Research Center for Intelligent Computing Platforms, Zhejiang Laboratory, Hangzhou 311100, China\\
$^{4}$E.A. Milne Centre for Astrophysics, University of Hull, Cottingham Road, Kingston-upon-Hull, HU6 7RX, UK\\
$^{5}$Centre of Excellence for Data Science, Artificial Intelligence and Modelling (DAIM), University of Hull, Cottingham Road, Kingston-upon-Hull, HU6 7RX, UK\\
$^{6}$Cahill Center for Astronomy and Astrophysics, MC 249-17 California Institute of Technology, Pasadena, CA 91125, USA\\
$^{7}$Owens Valley Radio Observatory, California Institute of Technology, 100 Leighton Lane, Big Pine, CA, 93513, USA\\
$^{8}$Department of Physics and Astronomy and the Center for Gravitational Waves and Cosmology, West Virginia University, Morgantown, WV 26506, USA
}

\date{Accepted XXX. Received YYY; in original form ZZZ}

\pubyear{2022}

\begin{document}
\label{firstpage}
\pagerange{\pageref{firstpage}--\pageref{lastpage}}
\maketitle

\begin{abstract}
Fast radio bursts (FRBs) are millisecond-timescale radio transients, the origins of which are predominantly extragalactic and likely involve highly magnetized compact objects. FRBs undergo multipath propagation, or scattering, from electron density fluctuations on sub-parsec scales in ionized gas along the line-of-sight. Scattering observations have located plasma structures within FRB host galaxies, probed Galactic and extragalactic turbulence, and constrained FRB redshifts. Scattering also inhibits FRB detection and biases the observed FRB population. We report the detection of scattering times from the repeating FRB~20190520B that vary by up to a factor of two or more on minutes to days-long timescales. In one notable case, the scattering time varied from $7.9\pm0.4$~ms to less than 3.1~ms ($95\%$~confidence) over 2.9 minutes at 1.45~GHz. The scattering times appear to be uncorrelated between bursts or with dispersion and rotation measure variations. Scattering variations are attributable to dynamic, inhomogeneous plasma in the circumsource medium, and analogous variations have been observed from the Crab pulsar. Under such circumstances, the frequency dependence of scattering can deviate from the typical power-law used to measure scattering. Similar variations may therefore be detectable from other FRBs, even those with inconspicuous scattering, providing a unique probe of small-scale processes within FRB environments. 
\end{abstract}

\begin{keywords}
transients: fast radio bursts -- stars:neutron -- stars: magnetars -- scattering -- plasmas
\end{keywords}    

\section{Introduction}

FRB 20190520B is only the second fast radio burst (FRB) localized to a dwarf galaxy and associated with a compact persistent radio source (PRS), {presumably} from a synchrotron nebula surrounding the source \citep{niu2022}. Its total line-of-sight (LOS) integrated electron density ($n_e$), or dispersion measure {${\rm DM} = \int_0^{z_{\rm h}} n_e(l) dl = 1205\pm4$ pc cm$^{-3}$}, is dominated by the host galaxy at redshift $z_{\rm h} = 0.241$, which contributes $\rm DM_h = 903^{+72}_{-111}$ pc cm$^{-3}$ (observer frame), at least three times the DM typically {inferred} for the host galaxies of non-localized FRBs {\citep{niu2022,luo2018,shin2022}}. Like some other repeating FRBs, FRB 20190520B shows extreme variations in rotation measure (RM), which are attributed to path-integrated magnetic field changes within its local environment \citep{feng2022,annathomas2022, dai2022}. 

FRB 20190520B also shows evidence of significant scattering, observed as both pulse broadening with a corresponding temporal delay $\tau$ (aka the scattering time), and scintillation with a corresponding frequency bandwidth $\Delta \nu_{\rm d}$. In \cite{ocker2022}, hereafter \citetalias{ocker2022}, we measured a mean scattering time $\bar{\tau}(1.41\ {\rm GHz}) = 10.9\pm1.5$ ms ($9.5\pm1.3$ ms at 1.45 GHz) and a mean scintillation bandwidth $\Delta \bar{\nu}_{\rm d}(1.41\ {\rm GHz})=0.21\pm0.01$ MHz ($0.23\pm0.01$ MHz at 1.45 GHz) for this source. Attributing $\bar{\tau}$ to the host galaxy and $\Delta \bar{\nu}_{\rm d}$ to the Milky Way constrained the mean scattering from the host galaxy to within 100 pc of the source.\\
\indent In this work we examine individual bursts from FRB 20190520B to characterize scattering variations near the FRB source. Unlike Galactic pulsar scattering, which even for the Crab pulsar varies slowly (longer than days to weeks; \citealt{mckee2018}), we find that the scattering time can vary significantly between bursts, indicating the presence of plasma inhomogeneities likely on sub-astronomical unit (au) scales within the circumsource medium (CSM). Section~\ref{sec:methods} describes the methods used to analyze burst spectra and constrain scattering. Results are presented in Section~\ref{sec:results} {and compared to other observations of the source in Section~\ref{sec:comparison}. Section~\ref{appendix:patches} explores a possible model for the plasma inhomogeneities that give rise to the scattering variations. Implications for the CSM and other FRB sources are discussed further in Section~\ref{sec:discussion}.}

\section{Methods}\label{sec:methods}
\begin{figure*}
    \centering
    \includegraphics[width=\textwidth]{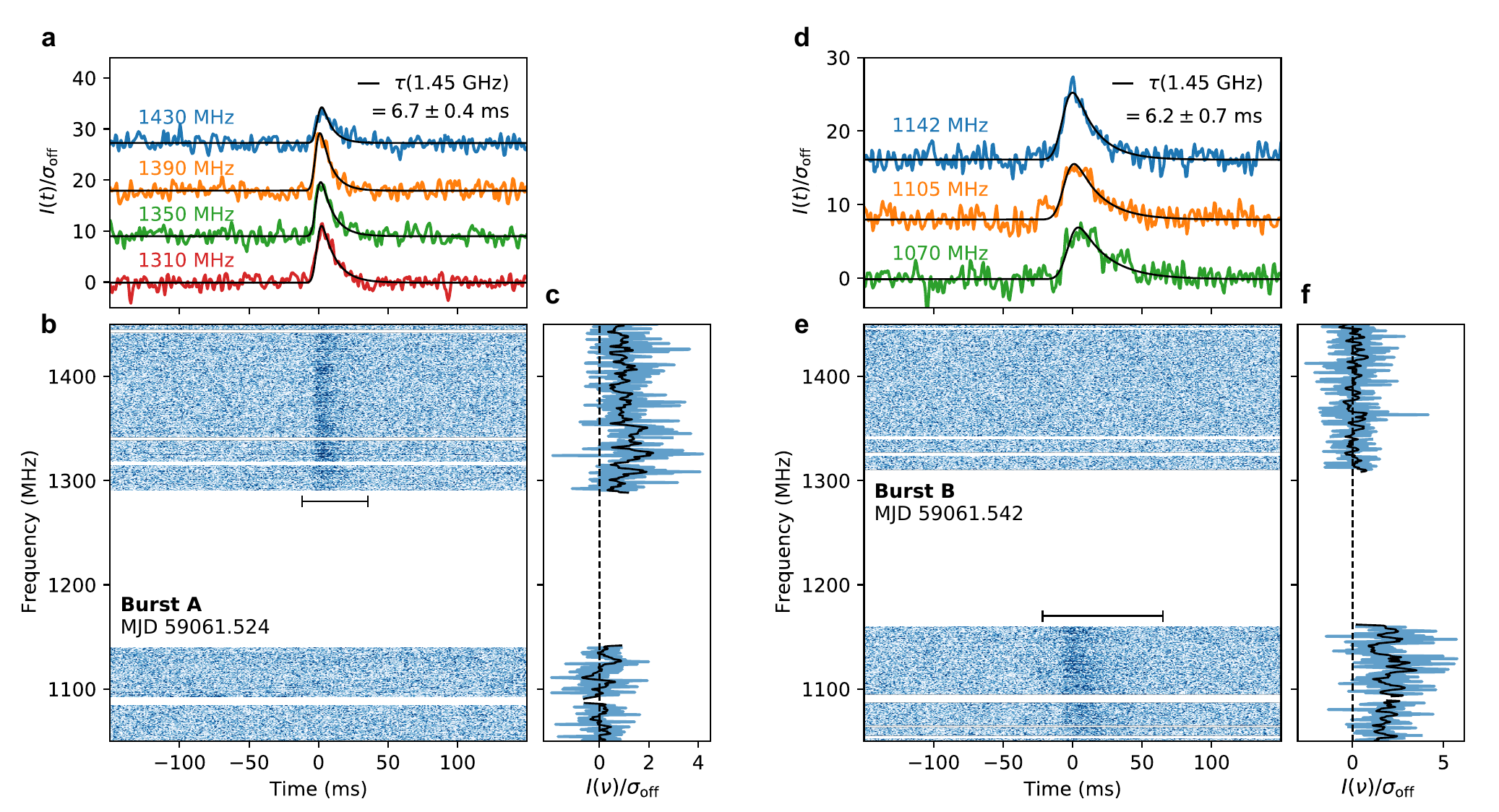}
    \caption{Two scattered bursts detected within 26 minutes. a) Frequency-averaged burst intensity vs. time in units of the signal $I(t)$ divided by the off-pulse noise $\sigma_{\rm off}$ and in four subbands centered on frequencies 1430 MHz (blue), 1390 MHz (orange), 1350 MHz (green), and 1310 MHz (red). Subbands are offset by an arbitrary amount for clarity. Black curves indicate the result of fitting a Gaussian pulse convolved with a one-sided exponential using a least squares fit for a Gaussian width fixed across frequency $\nu$ and a scattering time $\tau \propto \nu^{-4}$. The time resolution is 1.6 ms. b) Dynamic spectrum indicating burst intensity as a function of frequency vs. time. White bands are masked radio frequency interference (RFI). The horizontal black bar indicates the region used to calculate the time-averaged burst spectrum. c) Time-averaged burst spectrum vs. frequency $\nu$ in units of the signal-to-noise $I(\nu)/\sigma_{\rm off}$ (blue curve) and the spectrum smoothed with a 1 MHz-wide boxcar filter (black curve). d) - f) Same as a) - c) for a burst detected below 1200 MHz. The scattering time was fit using the same procedure applied to the three frequency subbands shown in panel d): 1075 MHz (green), 1125 MHz (orange), and 1150 MHz (blue).}
    \label{fig:burst_set1}
\end{figure*}

\begin{figure*}
    \centering
    \includegraphics[width=\textwidth]{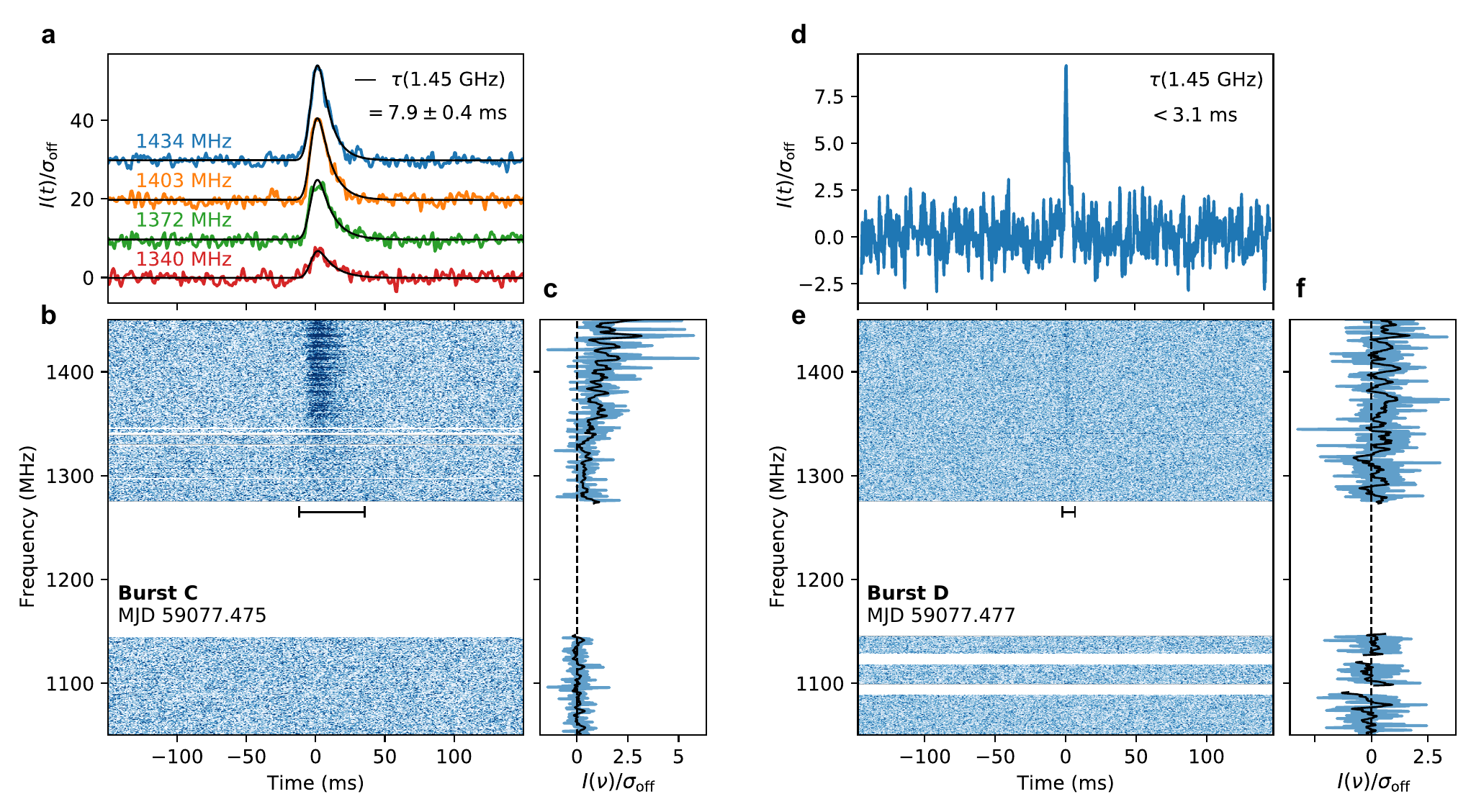}
    \caption{Two consecutive bursts with and without evidence of scattering. a) - c) Same as in Figure~\ref{fig:burst_set1} for a burst detected at MJD 59077.475, with a scattering time $\tau = 7.9\pm0.4$ ms at 1.45 GHz. d) - f) Same as a) - c) for a burst detected at MJD 59077.477. In this case, the peak S/N is too low to divide the burst into multiple frequency subbands and test for frequency-dependent scattering. The burst full-width-at-half-maximum implies a $95\%$ confidence upper limit on the scattering time $\tau < 3.1$ ms at 1.45 GHz.}
    \label{fig:burst_set2}
\end{figure*}

FRB 20190520B was initially detected in archival data from the Commensal Radio Astronomy FAST Survey (CRAFTS; \citealt{li18,nan11}). The burst sample considered in this paper is drawn from tracking observations of the FRB conducted at FAST between April and September 2020, which yielded 75 burst detections across 12 observing epochs in the 1.05 -- 1.45 GHz frequency band. These observations were discussed in \cite{niu2022}, and correspond to bursts P5 - P79 in the supplementary information of that paper (for reference, bursts A-D in Figures~\ref{fig:burst_set1}-\ref{fig:burst_set2} correspond to bursts P28, P34, P66, and P67). The same set of bursts was discussed in \citetalias{ocker2022}.

Bursts from FRB 20190520B show a range of morphologies, from burst intensities that are symmetric in time, to spectral islands that drift downward in frequency-time space (the ``sad-trombone"), and frequency-dependent temporal widths and intensity modulations that are attributable to scattering \citep{niu2022,ocker2022}. We have taken a number of steps throughout the analysis to mitigate confusion of intrinsic structure with scattering asymmetries, including the exclusion of bursts with multiple identifiable components and frequency-time drift from the analysis; the assessment of scattering models in multiple frequency subbands; and the statistical evaluation of burst asymmetries used in the skewness method described below.

\subsection{Initial Data Processing}

The data were initially recorded in filterbank format with a frequency resolution of 0.122 MHz and a sampling time of 49.5 $\mu$s. The data were subsequently smoothed to a temporal resolution of 1.57 ms using a 1D boxcar filter, except for burst D, for which a temporal resolution of 0.59 ms was used to obtain adequate sampling across the burst {due to its exceptionally narrow temporal width}. 

Two de-dispersion methods were explored, one that maximizes burst substructure \citep{hessels2019} and one that maximizes the burst {signal-to-noise ratio (S/N), defined as burst peak intensity divided by the root-mean-square (rms) of the off-burst noise} \citep{cordes2003}. While structure-optimization is generally favored for bursts that have multiple components and non-dispersive frequency-time drift, the scattering times of such bursts are highly ambiguous even after de-dispersion. We therefore removed bursts {identified by eye to have} non-dispersive frequency-time drift and/or multiple identifiable components (peak $\rm S/N \gtrsim 5$ when averaged across the entire 400 MHz band) and did not consider these bursts in subsequent analysis. For the remaining single-component bursts, we compared the structure-optimized DMs determined in \cite{niu2022} and S/N-maximized DMs, which were determined by calculating S/N for a range of trial DMs at 0.1 pc cm$^{-3}$ resolution. The peak and width of the resulting ambiguity function were used to determine the best-fit DM and error. There was minimal difference between the structure-optimized and S/N-maximized DMs for most of the single-component bursts in the sample. However, in some cases structure optimization misestimated the DMs of single-component bursts by failing to align the leading edge of intensity across all frequencies, which can result in an overestimated scattering time.\footnote{{Preliminary analysis suggests that structure optimization may appear to misestimate DMs when the intensity varies enough within a burst that brighter components of the burst get overweighted with respect to fainter components. Multi-path propagation may also play a role here, as different paths may have slightly different DMs.}} We therefore use the S/N-maximized DMs in subsequent analysis.
All dynamic spectra were individually examined to affirm that the leading edge of intensity was aligned across the burst bandwidth, before proceeding with the scattering analysis.\\
\indent In most cases, burst intensity is concentrated above 1.3 GHz. {We define the burst bandwidth using the minimum and maximum frequencies where the time-averaged burst spectrum has a $\rm S/N > 2$, for a fixed time window of 300 ms around each burst. The central frequency is taken to be the mid-frequency of the burst bandwidth (without any weighting by intensity).}
The average central frequency of the burst sample is 1.35 GHz. Data from 1.16-1.29 GHz were masked for most bursts due to radio frequency interference (RFI).

\subsection{Empirical Burst Widths}

We measure the total, empirical width of each burst using the ACF of the burst intensity averaged over the burst's entire spectral bandwidth, $\langle I(t) I(t + \delta t) \rangle$. The ACF error at a lag $k$ is $\sqrt{(1/n)\times (1+2\sum_{m=1}^{k-1}r_m^2)}$, where $n$ is the length of the time series and $r_m$ is the autocorrelation at lag $m$ {\citep{priestley1981}}. The burst full-width-at-half-maximum (FWHM) is estimated using the half-width-at-half-maximum (HWHM) of the ACF, ${\rm FWHM} = \sqrt{2} \times \rm HWHM_{ACF}$ (calculated after removal of the noise spike at zero lag). For a Gaussian burst, this is equivalent to the FWHM that would be derived directly from the pulse shape. In general, ${\rm FWHM} \approx \sqrt{W_i^2 + W_{\rm PBF}^2}$, where $W_i$ is the intrinsic burst width and $W_{\rm PBF}$ is the width of the {pulse broadening function (PBF)}. 

\subsection{Burst Scattering Times}

A canonical, robust scattering measurement generally requires that the burst intensity be asymmetric in time, with an extended scattering tail that increases at lower observing frequencies. Accurate identification of scattering thus requires an assessment not only of the pulse profile in time, but also the evolution of that profile over frequency, which in turn requires precise de-dispersion. For fitting purposes, the burst profile (intensity vs. time) is assumed to consist of a Gaussian pulse convolved with a one-sided exponential PBF, with a $1/e$ delay $\tau$ that scales with observing frequency $\nu$ as $\tau \propto \nu^{-4}$. The Gaussian pulse is assumed constant in frequency $\nu$, while the scattering time $\tau$, the $1/e$ time of the PBF, evolves as $\tau \propto \nu^{-4}$. Each burst was divided into multiple frequency subbands before averaging over frequency to obtain the temporal burst profile as a function of frequency. The scattering time and Gaussian width were then fit by minimizing the $\chi^2$ statistic, and the burst amplitude was left as a free parameter that varied between subbands. While PBFs discerned from pulsar observations can be non-exponential and intrinsic widths can vary with frequency, our simple approach is sufficient for the data in hand.

Scattering times are only reported for bursts that satisfy two main criteria: 1) A combination of sufficient S/N and burst bandwidth -- in practice, a $\rm S/N \gtrsim 5$ in at least 2 frequency subbands, {where a given subband typically needed to be $> 20$ MHz wide to give the required S/N}; and 2) there is a global minimum in $\chi^2$, {a reduced $\bar{\chi}^2\approx 1$}, and $\tau$ has a fractional error $<30\%$. We refer to bursts that fit these critera as Set 1. Bursts that do not meet these criteria are called Set 2. Set 2 contains both low S/N bursts that do not meet criterion (1), and high S/N bursts that do not meet criterion (2). {Scattering may still be relevant to Set 2 bursts because larger scattering can reduce burst S/N, to the point where criterion (1) is no longer met, and because inhomogeneities in the CSM may cause non-exponential PBFs (see Section~\ref{sec:discussion} for further discussion). The methods used to assess these effects are described in the following two sections.}

\subsection{The Skewness Test}

To constrain the presence of scattering for bursts in Set 2, we develop a two-part metric based on the skewness function \citep{stinebring81}. The skewness function quantifies the degree and direction of asymmetry in a burst of intensity $I(t)$, and is given by
\begin{equation}\label{eq:skewness}
    \kappa(\delta t) = \frac{\langle I^2(t) I(t+\delta t)\rangle - \langle I(t)I^2(t+\delta t)\rangle}{\langle I(t)\rangle^3},
\end{equation}
where brackets denote time averages and $\delta t$ is a given time lag. Typically the skewness function is normalized using the third moment $\langle I^3(t) \rangle$, but this normalization yields a strong S/N dependence that renders large errors for many bursts in our sample. We mitigate this effect by normalizing with the mean $\langle I(t) \rangle^3$. For an asymmetric pulse, the skewness function is antisymmetric in $\delta t$, and maximizes at an amplitude $\kappa_{\rm max}$ and a lag $\delta t_{\rm max}$. When calculating $\kappa_{\rm max}$ and $\delta t_{\rm max}$ we only consider lags less than twice the burst width inferred from the ACF.

The two-part skewness test assesses both the sign of $\delta t_{\rm max}$ and the amplitude $\kappa_{\rm max}$. For an exponential PBF, $\kappa(\delta t)$ maximizes at $\delta t_{\rm max} = \tau \rm ln2$, and $\kappa_{\rm max}$ increases with respect to $\tau$. For a Gaussian pulse convolved with an exponential PBF, $\delta t_{\rm max}/ \rm ln2 > \tau$. Noise can induce both positive and negative temporal asymmetries. For high S/N bursts this effect is negligible and the sign of $\delta t_{\rm max}$ for an individual burst provides one piece of evidence for scattering. A sample of noisy, intrinsically symmetric bursts will have equal probability of $\delta t_{\rm max}$ being positive or negative, but a sample of noisy, scattered bursts will preferentially have $\delta t_{\rm max} > 0$. One could also argue that intrinsically asymmetric bursts will not preferentially be biased towards positive temporal asymmetries, {depending on the emission mechanism (which remains highly uncertain)}. The distribution of $\delta t_{\rm max}$ for a sample of independent bursts is thus also used to assess the presence of scattering. 

The second part of the skewness test assesses the amplitude $\kappa_{\rm max}$. When the S/N of a given burst is high (the exact S/N threshold depends on the burst width; see Appendix~\ref{appendix:skew}), the amplitude of the maximum skewness $\kappa_{\rm max}$ is compared to the maximum skewness of an exponential PBF with the same total width as the observed burst. The resulting ratio of skewness amplitudes is then compared to the ratio that would be obtained for a Gaussian burst, in order to determine whether the observed skewness is consistent or inconsistent with scattering to a given statistical confidence level. The full procedure for assessing the skewness amplitude $\kappa_{\rm max}$ is described in Appendix~\ref{appendix:skew}.

\subsection{Mean Scattering Times from Fourier Domain Stacking}

In \citetalias{ocker2022} we demonstrated that stacking bursts' temporal profiles in the Fourier domain can be used to infer an average scattering time. This method has the advantage of mitigating shifts in burst arrival times both across the frequency band of a single burst and when stacking different bursts. Here we employ an identical routine to compare the average scattering times of bursts in Sets 1 and 2, {in order to test whether Set 2 bursts have an average scattering time larger than the scattering times in Set 1 (as may be expected if weaker bursts are more scattered).} Burst temporal profiles were obtained by averaging each burst over two frequency subbands, 1.29-1.37 GHz, and 1.37-1.45 GHz. The power spectra (equivalent to the squared magnitude of the fast Fourier transform) of all burst profiles falling within a given subband were then stacked to compute an average power spectrum for each frequency subband. In \citetalias{ocker2022}, an additional frequency subband from 1.05-1.25 GHz was used, but the number of bursts falling within this subband is too small to compute an average power spectrum for this subband from Sets 1 and 2 separately. The average power spectrum was then fit with the canonical scattering model, where the spectrum consists of the product of Gaussian and PBF contributions. The error in $\tau$ inferred from this method includes contributions from the rms fluctuations of the individual power spectra about the mean spectrum, and from the rms residuals between the mean power spectrum and the fitted model. A complete description of the stacking method is provided in \citetalias{ocker2022}.

\section{Analysis \& Results}\label{sec:results}
\subsection{Set 1 Bursts: Measurement of Scattering Variations}
Bursts A-C in Figure~\ref{fig:burst_set1} and Figure~\ref{fig:burst_set2}a show examples of bursts in Set 1, which have best-fit scattering times of $6.7\pm0.4$ ms, $6.2\pm0.7$ ms, and $7.9\pm0.4$ ms at 1.45 GHz for bursts A, B, and C, respectively.
{Set 1 also contains significant scattering measurements for thirteen other bursts (Table~\ref{tab:tab1}), which we compare to bursts in Set 2 below.} 

\begin{table}
\centering
\begin{tabular}{c|c|c|c}
     & $\tau$ & Gaussian FWHM & DM \\
    MJD & (ms at 1.45 GHz) & (ms) &  (pc cm$^{-3}$) \\
    \hline \hline
    58991.68687 & $6.5\pm0.9$ & $5.9\pm0.7$ & $1210\pm5$ \\
    58991.70463 & $6.7 \pm 1.4$ & $5.7\pm2.2$ & $1210\pm 4$ \\
    58991.71769 & $7.4\pm0.3$ & $7.5\pm0.2$ & $1219\pm 5$ \\
    58991.71788 & $6.1\pm 1.5$ & $5.7\pm1.5$ & $1222\pm 4$ \\
    58991.71822 & $8.5\pm0.2$ & $10.6\pm0.2$ & $1211\pm 7$ \\
    59060.48447 & $7.9\pm 0.3$ & $9.7\pm0.5$ & $1187\pm11$ \\
    59060.50785 & $6.9\pm0.3$ & $4.2\pm0.5$ & $1196\pm 13$ \\
    59060.52596 & $7.0\pm 0.3$ & $6.8\pm0.4$ & $1205\pm10$ \\
    59061.52434\textsuperscript{A} & $6.7\pm 0.4$ & $5.4\pm0.4$ & $1196\pm 9$ \\
    59061.54182\textsuperscript{B} & $6.2\pm 0.7$ & $12.2\pm1.2$ & $1210\pm 5$ \\
    59067.50989 & $6.9\pm 0.5$ & $3.5\pm0.5$ & $1213\pm5$ \\
    59067.53524 & $5.9\pm 0.4$ & $6.8\pm0.5$ & $1202\pm9$ \\
    59069.51499 & $7.6\pm0.5$ & $5.4\pm0.2$ & $1210\pm 7$ \\
    59077.46629 & $6.9 \pm 0.4$ & $7.1\pm0.5$ & $1190\pm 4$ \\
    59077.46990 & $9.1\pm 0.7$ & $4.5\pm0.2$ & $1180\pm 10$ \\
    59077.47533\textsuperscript{C} & $7.9\pm 0.4$ & $8.9\pm0.5$ & $1217\pm 11$ \\ 
    \hline 
    59077.47744\textsuperscript{D} & $< 3.1$ & $2.9\pm0.1$ & $1197\pm3$ 
\end{tabular}
\caption{\label{tab:tab1}Scattering times for bursts in Set 1. Burst arrival times are quoted in modified Julian date (MJD) to a precision of about one second, and are referenced to the Solar System barycentre at 1.5 GHz. Scattering times $\tau$ were measured using a 2D fitting routine that assumes $\tau \propto \nu^{-4}$, and are referenced to 1.45 GHz by the same assumption. The fitting also assumes the Gaussian FWHM is constant across frequency. DMs shown maximize the S/N. {Bursts A-D are indicated with superscripts.}}
\end{table}

\begin{figure}
    \centering
    \includegraphics[width=80mm]{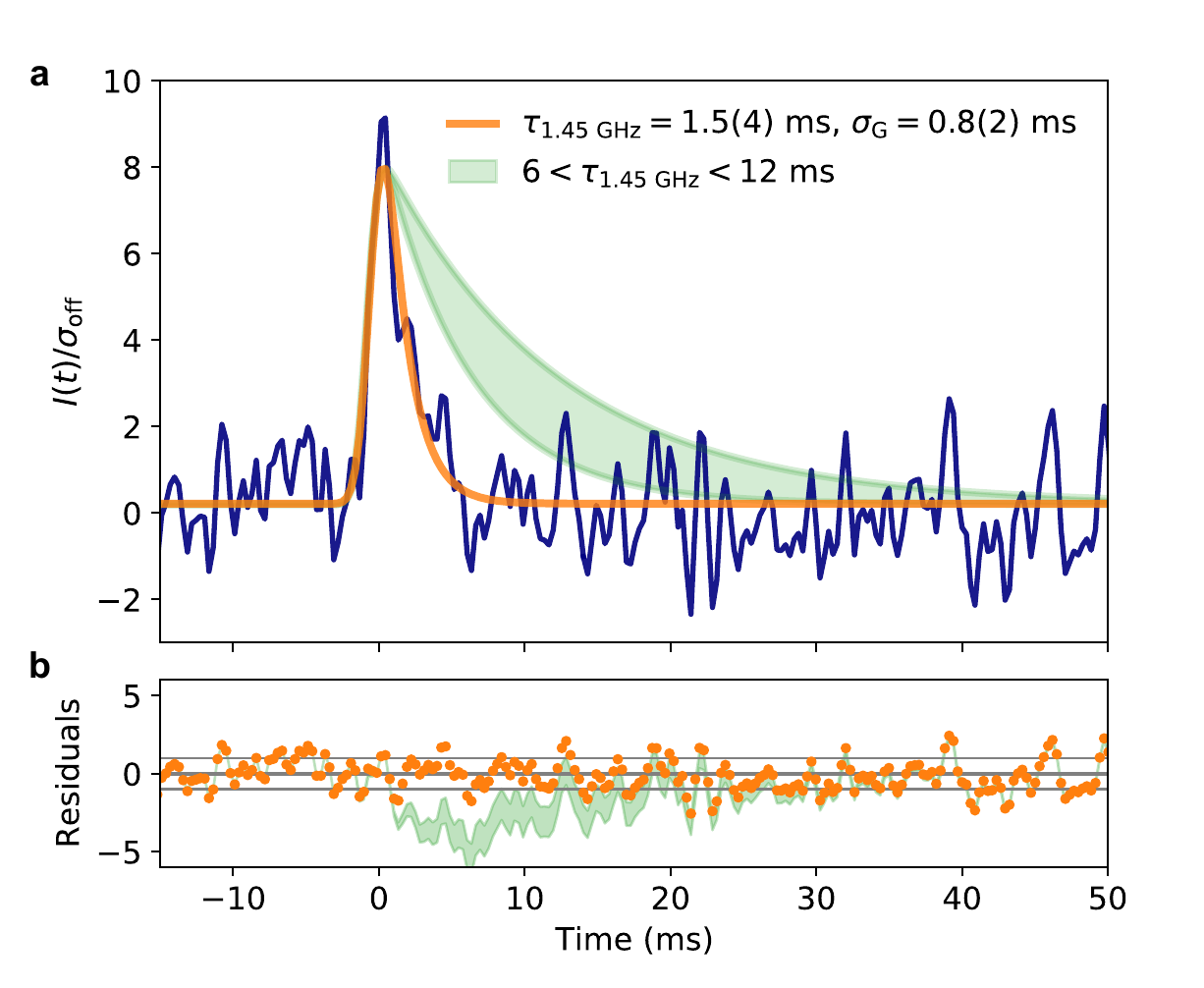}
    \caption{Comparison of the narrowest burst with average scattering. a) The dark blue curve shows the frequency-averaged burst intensity vs. time in signal-to-noise units $I(t)/\sigma_{\rm off}$ for burst D, with a time resolution of 0.6 ms. The orange curve shows the result of fitting a Gaussian pulse convolved with an exponential PBF to the burst intensity, which yields a scattering time $\tau=1.5 \pm 0.4$ ms and a Gaussian standard deviation $\sigma_G=0.8 \pm 0.2$ ms. The reference frequency for $\tau$ is taken to be the highest frequency at which the burst is detected, 1.45 GHz. {The filled green region demonstrates the range of scattering tails that would correspond to the same Gaussian width and the range of scattering times measured for other bursts in the sample, normalized to the same peak intensity as the orange model.} b) Residuals between the measured burst intensity and the orange and green models shown in panel a). Residuals of $\pm 1$ are indicated by the grey horizontal lines.} 
    \label{fig:62_modelcomparison}
\end{figure}

{Burst D (Figure~\ref{fig:burst_set2}b) stands out as having a much shorter scattering time than bursts in Set 1.
It was detected only 2.9 minutes after burst C, with a scattering time that is at least a factor of two smaller.} The ACF of the frequency-averaged burst profile yields an empirical measurement of the burst FWHM, $W = 2.9 \pm 0.1$ ms; the contribution of intra-channel dispersion smearing to the burst width is less than $1\%$. The burst peak $\rm S/N = 9.1$ is too small to perform a least squares fit for scattering in both frequency and time. Figure~\ref{fig:62_modelcomparison} shows the results of fitting the 1D burst profile with a Gaussian pulse convolved with an exponential PBF, which yields $\tau(1.45\ {\rm GHz}) = 1.5 \pm 0.4$ ms and a standard deviation $\sigma_G = 0.8 \pm 0.2$ ms with a reduced $\bar{\chi}^2 = 1.2$. For the same Gaussian width, scattering times between 6 and 12 ms at 1.45 GHz (the approximate range of $\tau$ across the entire burst sample) would yield much larger temporal widths than observed from the burst profile. Fitting a symmetric Gaussian pulse to the burst yields $\sigma_G = 1.3\pm0.1$ ms with $\bar{\chi}^2 = 1.4$, and hence is not preferred over the exponential model. As the peak S/N is too small to assess the frequency dependence of the burst width, we place a $95\%$ confidence upper limit on the scattering time of $\tau < 3.1$ ms at 1.45 GHz, based on the empirical burst width measured from the ACF. The scattering reference frequency is (conservatively) taken to be the highest frequency at which the burst is detected. {The DM of burst D ($1197\pm3$ pc cm$^{-3}$) is marginally different from that of burst C ($\rm DM = 1217 \pm 11$ pc cm$^{-3}$).} The $\tau$ upper limit for burst D is at least two times smaller than the scattering times measured for bursts A-C and the other bursts with individual scattering measurements, all of which are shown in Figure~\ref{fig:w_vs_tau}. 

\subsection{Comparison Between Sets 1 and 2 Bursts}
\begin{figure}
\centering
  \centering
  \includegraphics[width=0.4\textwidth]{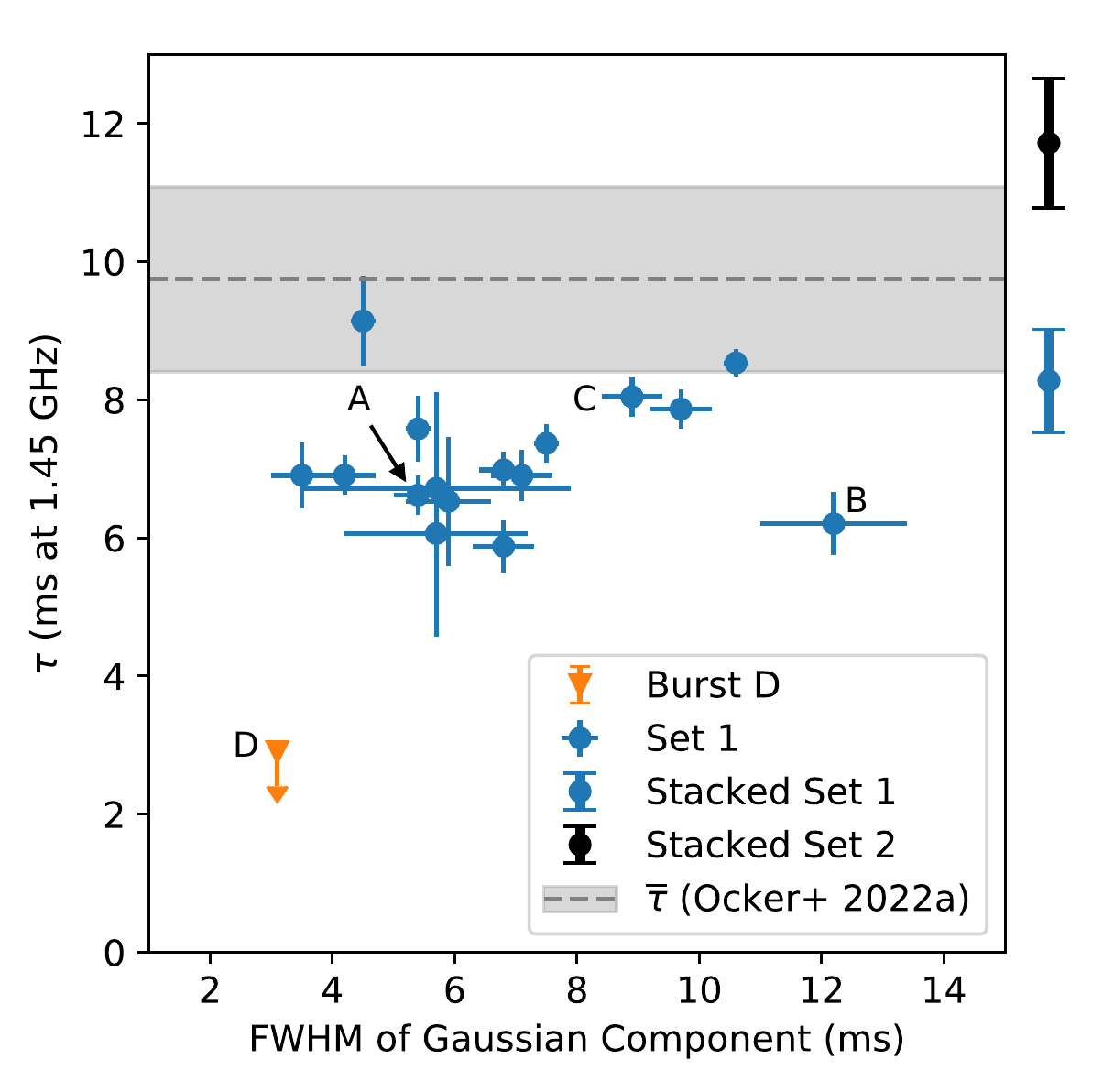}
  \caption{Burst scattering times and Gaussian widths. Blue points and errors correspond to the best-fit values and $68\%$ confidence intervals for the scattering time $\tau$ in milliseconds at 1.45 GHz (top of the observing band) and the FWHM of the Gaussian burst component, fit using the same procedure as for bursts A-D. Sets 1 and 2 refer to bursts with and without significant least squares fits for $\tau$ and the Gaussian width, respectively. Also shown in orange is the upper limit on $\tau$ for burst D, the narrowest burst in the sample. The grey dashed line and shaded region correspond to the mean and standard deviation of $\tau$ inferred from stacking Sets 1 and 2 together in the Fourier domain \citep{ocker2022}. The blue and black capped lines indicate the mean of $\tau$ and its standard deviation, inferred from applying the same stacking method to Sets 1 and 2, respectively (Methods).}
  \label{fig:w_vs_tau}
\end{figure}

\begin{figure}
\centering 
  \includegraphics[width=0.4\textwidth]{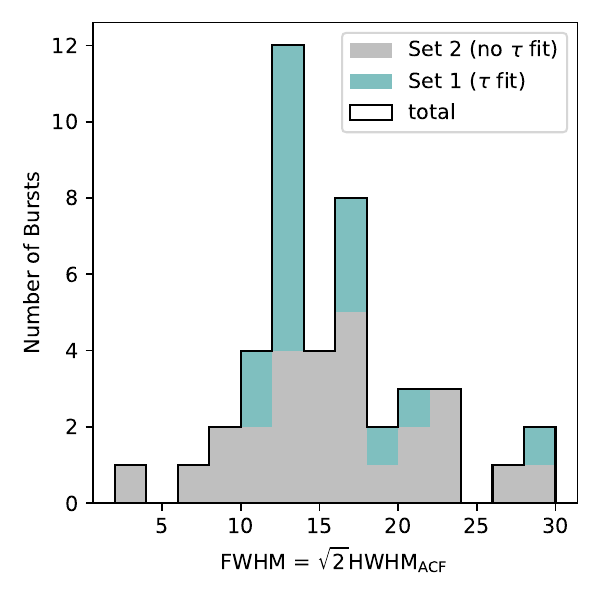}
  \caption{Distribution of total burst widths. The burst FWHM is defined as $\sqrt{2}\times$ HWHM of a burst's ACF, calculated from the burst profile integrated across the entire frequency band, 1.05 -- 1.45 GHz (Methods). Full-widths for bursts in Set 1 (which have individual scattering measurements) are shown in green, and burst widths in Set 2 (no individual scattering measurements) are shown in grey. The average central frequency of the burst emission is 1350 MHz (Methods), and the total widths shown here are consistent with including the contributions of both scattering and intrinsic structure.}
  \label{fig:widthhist}
\end{figure}

The scattering times shown for individual bursts in Figure~\ref{fig:w_vs_tau} represent cases with both sufficient S/N and spectral bandwidth to perform a least squares fit that yields significant scattering measurements (these bursts constitute Set 1; see Section~\ref{sec:methods}).
From these  bursts alone, one would infer that the mean scattering time is $6.9\pm1.0$ ms at 1.45 GHz, and that $\tau$ can fluctuate by at least a factor of two between bursts. However, bursts in Set 1 only constitute a fraction of the bursts observed, {and the scattering variations measured for Set 1 are not necessarily representative of the full range of scattering that may occur}. Set 2 contains 32 other bursts, five of which are high S/N bursts that do not show the frequency dependence assumed in the canonical scattering model, either because they are inconsistent with any frequency-dependent temporal broadening or because their temporal widths decrease at lower observing frequencies. The rest of the Set 2 bursts have too low $\rm S/N$ to evaluate the scattering model on an individual burst basis.\\
\indent In \citetalias{ocker2022}, we used Fourier domain stacking of bursts' temporal profiles (Section~\ref{sec:methods}) to determine a mean scattering time $\bar{\tau}(1.45\ {\rm GHz}) = 9.5\pm1.3$ ms for the same burst sample analyzed here. This mean scattering time was obtained using both high and low S/N bursts for which scattering can and cannot be measured individually, and is larger than most of the scattering times shown for Set 1 in Figure~\ref{fig:w_vs_tau}. {This difference is partially due to a trade-off between intrinsic width and scattering: In general, $\tau$ can only be fit using the canonical scattering model when $\tau$ is greater than the intrinsic width in at least part of the frequency band.} 
Figure~\ref{fig:widthhist} shows the distribution of total widths measured for bursts in Sets 1 and 2. Set 2 does contain more bursts with larger widths than Set 1, but a two-sided Kolmogorov-Smirnov test between the widths of Sets 1 and 2 bursts yields a p-value $=0.4$, indicating that the total widths of the two burst sets are statistically consistent with being drawn from the same distribution. We also find no evidence of a strong correlation between burst total width and S/N in either burst set.\\
\indent In order to assess whether the burst widths in Set 2 do include contributions from scattering, rather than simply having larger intrinsic widths, we examine both the skewness functions of the bursts and re-perform the stacking analysis on Sets 1 and 2 separately.\\
\indent Using the skewness test, we find that two bursts in Set 2 have skewness functions with significant evidence of scattering, based on both their skewness amplitudes and sign of  $\delta t_{\rm max}$ {(see Figure~\ref{fig:maxskew3746} in Appendix~\ref{appendix:skew})}. The skewness test was inconclusive for most of the bursts in Set 2 because their S/N is too low to assess whether their maximum skewness amplitudes are consistent or inconsistent with scattering. Nonetheless, there are $8\times$ more bursts with positive $\delta t_{\rm max}$ than negative $\delta t_{\rm max}$ in Set 2, demonstrating that the sample of bursts in Set 2 is largely dominated by positive-handed temporal asymmetries. The distribution of $\delta t_{\rm max}$ for Set 2 is thus inconsistent with a population of intrinsically symmetric bursts with asymmetries contributed by noise alone. (For reference, all bursts in Set 1 have $\delta t_{\rm max} > 0$.) We therefore conclude that bursts in Set 2 are overwhelmingly asymmetric and skewed to positive lags. While we have excluded bursts with identifiable sad-trombone drift from Set 2, we note that even unresolved drifting or an imprecise DM would not necessarily cause bursts to be preferentially skewed to positive lags.\\
\indent The skewness test indicates that scattering is likely present in Set 2 bursts. We therefore apply the same Fourier domain stacking analysis used in \citetalias{ocker2022} to Sets 1 and 2 separately, in order to determine whether the scattering in Set 2 is significantly different from that in Set 1. The mean and standard deviation of $\tau$ inferred from this analysis is shown in Figure~\ref{fig:w_vs_tau} for both Sets 1 and 2. For Set 1, the stacking analysis yields $\bar{\tau}(1.45\ {\rm GHz}) = 8.0\pm0.7$ ms, whereas for Set 2, the stacking analysis yields $\bar{\tau}(1.45\ {\rm GHz}) = 11.3\pm0.9$ ms. The mean of these values is consistent with the result presented in \citetalias{ocker2022}, which found $\bar{\tau}(1.45\ {\rm GHz}) = 9.5\pm1.3$ using both Sets 1 and 2. {The mean scattering time inferred for Set 1 from the stacking method is about 1 ms larger than the mean calculated by directly averaging the scattering times of individual bursts in Set 1, although the two methods give results that are technically consistent within one standard deviation. This comparison suggests that the stacking method may overestimate the mean scattering time by an amount comparable to the inferred uncertainties.} Simply comparing the mean scattering times for Sets 1 and 2 confirms that $\tau$ varies by at least $40\%$ across the burst sample, but comparing the scattering times of individual bursts in Set 1 to the mean scattering time of Set 2 suggests that $\tau$ can vary by up to $100\%$ or more at 1.45 GHz. 

\subsection{{Summary of Key Results}}
\indent The difference in $\tau$ between bursts C and D suggests that $\tau$ can vary significantly over a timescale as rapid as 2.9 minutes, and differences in $\tau$ are also seen between bursts detected on different days (Table~\ref{tab:tab1}). A significant change in $\tau$ over 2.9 minutes suggests that the length scale over which this change occurs is at most $c\Delta t \sim 0.4$ au, where $c$ is the speed of light. This scale is equivalent to an upper limit on the transverse offset between the two burst LOSs, which trace regions of significantly different scattering strength. This 0.4 au upper limit on the size scale is extremely conservative, given that the actual size scale is probably related to the relative velocity of the source $v \ll c$ (where $v$ is not known \textit{a priori)}. For typical pulsar velocities $\sim 100$ km/s \citep{verbunt2017} the size scale would be as small as thousands of kilometers.\\
\indent {Applying the skewness test to Set 2 demonstrates that scattering is likely present in these bursts, even though they do not meet our criteria for inferring burst scattering times using the canonical scattering model. Stacking Set 2 bursts in the Fourier domain yields a mean scattering time that is about $40-100\%$ larger than the scattering times of bursts in Set 1, suggesting that the scattering of FRB 20190520B can fluctuate more than we infer from Set 1 bursts alone.} \\ 
\indent We have not found any significant evidence of correlations between the scattering times of Set 1 bursts, or any significant evidence of secular trends over time. We also find no obvious evidence for a correlation between $\tau$ and DM in Set 1. There are apparent DM fluctuations $\approx 5-10$ pc cm$^{-3}$ between bursts, but these fluctuations are comparable to the measurement errors and may result from variations in burst structure. 
DM variations are expected at some level because variations in $\tau$ and RM are detected, {and independent study of bursts detected at Green Bank Telescope and Parkes Telescope suggests that there are burst-to-burst variations in DM, albeit without a significant long-term (months to years-long) trend} \citep{annathomas2022}. Future studies should continue to test for correlations in $\tau$ and DM between bursts, given that the sample of scattering times in Set 1 is sparse compared to the total number of bursts detected.

\section{Comparison to Other Observations of the Source}\label{sec:comparison}
Independent observations of FRB 20190520B have associated the FRB with a persistent radio source \citep{niu2022} and RM variations over days to months have been detected from the FRB at frequencies above 2 GHz \citep{annathomas2022,dai2022}. Previous searches for RM in the same data set discussed in this work have yielded non-detections, with an upper limit of $20\%$ on the degree of linear polarization \citep{feng2022,niu2022}. 
The degree of linear polarization increases substantially at higher frequencies \citep{feng2022,annathomas2022,dai2022}, suggesting that the non-detection of RM between 1.05 -- 1.45 GHz is related to multi-path scattering that reduces the degree of linear polarization \citep{beniamini2022,feng2022}. However, there is no empirical evidence of a direct correlation between the scattering and RM variations, as these phenomena are observed at distinct radio frequencies. {Moreover, the large difference in timescales over which the scattering and RM variations are observed (minutes for the former, and days to months for the latter) suggests that these phenomena may arise from separate screens in the CSM.} \\
\indent All of these observations indicate a dynamic, multi-phase source environment. Scattering variations, in particular, imply fluctuations in weakly or non-relativistic, thermal ionized gas along the LOS. In the following section we consider one possible model that explains such fluctuations as a distribution of ionized cloudlets, or ``patches," that are slightly offset from the direct LOS.

\section{Scattering from Discrete Patches}\label{appendix:patches}

Here we give an example of a physical model that explains scattering variations in terms of discrete patches distributed near the source. This patch model will be expanded upon in a future paper. This framework can be extended to a range of physical scenarios in which the CSM is non-uniform. 

Consider a rotating emission beam whose luminosity is highly intermittent. The burst emission has a duration $\Delta t_e$ and a beam width $\Delta \theta_b$. The spin period of the beam is $P_s$. The emission beam rotates across a region of depth $L$, containing scattering patches of radius $r_c$ and total transverse size $\Delta x \equiv 2r_c$ at typical separations $\Delta l$. A scattering patch is located at a distance $d_{\rm sl}$ from the source, and a distance $d_{\rm lo}$ from the observer. The source-to-observer distance $d_{\rm so}$ and lens-to-observer distance $d_{\rm lo}$ are both much larger than $d_{\rm sl}$.

In this model, a single burst would encounter a small number of patches. 
For simplicity, we assume here just one patch is illuminated. The number density of patches is $n_l \sim (\Delta l)^{-3}$. The mean free path for encountering a patch is $l_{\rm mfp} = 1/(\pi n_l r_c^2)$. In order for a burst to encounter a single patch, $l_{\rm mfp} \lesssim L$, implying an upper limit on the patch number density $n_l \lesssim 1/(\pi r_c^2 L)$. The total number of patches in a spherical volume surrounding the source is then $N_l \lesssim (4/3)(L/r_c)^2$. 

There are two main constraints on the beam size $\Delta \theta_b$: It must be large enough to fully illuminate a patch, implying $\Delta \theta_b d_{\rm sl} \sim \Delta \theta_b L \gtrsim \Delta x$, and it must be small enough that only one patch is illuminated, implying $\Delta \theta_b n_l L^3/3 \sim 1$. {Taking $\Delta l$ to be a} multiple $m_l$ of the patch size, we then have $\Delta x/L \lesssim \Delta \theta_b \lesssim (3/L^2 n_l) \sim 3(m_l\Delta x/L)^3$, and $\Delta x/L \gtrsim \sqrt{1/3m_l^3}$. The beam size is thus
\begin{equation}
    \Delta \theta_b \lesssim (1/3m_l^3)^{1/2} \approx 6\times10^{-4} (m_l/100)^{-3/2},
\end{equation}
{where the fiducial value $m_l = 100$ corresponds to patches that are neither tightly packed nor extremely spread out.} With relativistic beaming at a Lorentz factor $\gamma$, $\Delta \theta_b \gtrsim 1/\gamma$, implying $\gamma \lesssim 1700(m_l/100)^{3/2}$. A larger separation between patches increases the upper bound on the Lorentz factor. Lorentz factors $\sim 10^3 - 10^5$ have been inferred for radio pulsars \citep{rudermansutherland75}, including $\sim 10^4$ for Crab giant pulses \citep{bij2021}, which provides one possible metric for comparing the emission mechanisms of FRBs and giant pulses.\\
\indent The beam duration $\Delta t_e$ must also be short enough that at most one patch is illuminated per spin period $P_s$. Assuming that the interval between bursts is much larger than $P_s$, we thus have $\Delta t_e \times (2\pi/P_s) \lesssim \Delta l / L$, or $\Delta t_e \lesssim \Delta l P_s / 2\pi L = m_l \Delta x P_s / 2\pi L$. For $P_s$ in seconds the emission duration is then 
\begin{equation}
    \Delta t_e \lesssim 8\ {\rm ms} \times \bigg(\frac{m_l}{100}\bigg) \bigg(\frac{\Delta x}{100\ \rm au}\bigg) \bigg(\frac{P_s}{1\ {\rm s}}\bigg) \bigg(\frac{1\ \rm pc}{L}\bigg).
\end{equation}
The narrowest burst we detect is $2.9\pm0.1$ ms wide, which points to either smaller $m_l$, $\Delta x$, or $P_s$, larger $L$, or some combination of the above. Nonetheless, emission durations on the order of milliseconds are entirely consistent with patches $\sim$tens of au in transverse size distributed within a $\sim 1$ pc wide region around the source. Each patch contributes a ${\rm DM} \sim 2 n_e r_c \approx 4.8\times 10^{-4}\ {\rm pc\ cm^{-3}} (r_c/50\ {\rm au})n_e$. Even for a density $\gg 1$ cm$^{-3}$, this DM would be extremely small compared to the total DM of FRB 20190520B, which may explain why we do not detect any obvious temporal correlations between the observed scattering and DM.

\section{Summary \& Discussion}\label{sec:discussion}
{We find that scattering times vary between individual bursts from FRB 20190520B. In one case, the scattering time varies by over a factor of two between two consecutive bursts detected 2.9 minutes apart. Such a rapid variation likely arises from plasma inhomogeneities within a parsec of the source on sub-au transverse spatial scales. There is no significant evidence for correlations or trends in the scattering times of individual bursts, or in correlations between scattering and apparent DM variations. These conclusions are ultimately limited by the sparseness of bursts that fit the canonical scattering model (Set 1). We present a methodology based on skewness that can be used in future studies to assess the presence of scattering, even for bursts that do not fit the canonical scattering model. Applying this methodology to Set 2 bursts indicates that scattering is likely present in many of these bursts, even though their individual scattering times cannot be inferred by traditional methods. Subsequent stacking of Set 2 burst profiles yields a mean scattering time $\bar{\tau}(1.45\ {\rm GHz}) = 11.3\pm0.9$ ms that is about $3\pm1$ ms larger than the mean scattering of bursts in Set 1.}\\
\indent {One possible model that can explain the observed scattering variations is a distribution of discrete patches of plasma in the CSM.} Conservation of scattered burst flux occurs only for a very wide screen with homogeneous scattering properties. However, a patchy CSM will cause dilution of burst flux in a manner that would likely correlate with scattering \citep{cl_trunc_screen2001}. Patches could also be regions of significantly less scattering than the surrounding volume, and in this case the flux would be diluted except for LOS that pass through the patches.  This effect may be difficult to identify in practice, given the large flux variability seen in FRBs for which scatter-broadening appears to be minimal (e.g. FRB 20121102A; \citealt{hessels2019,li2021}). For FRB 20190520B, {we find that Set 2, which includes many low S/N bursts, has a larger mean scattering time than Set 1. However, further assessment of both Set 1 and Set 2 bursts does not yield any} significant evidence of a correlation between burst total width and S/N, which would be one indicator of flux dilution from scattering (barring intrinsic flux variations, which are not accounted for). Refraction may also be relevant in the CSM. \\
\indent Analogous scattering variations have been observed from the Crab pulsar and are induced in its supernova remnant \citep{lyne75,backer2000,lyne2001,mckee2018}. Variations in the diffractive scattering time $\tau$ have been observed down to a resolution of 15 days over 30 years of archival data, and show a positive correlation with DM fluctuations $\lesssim 0.05$ pc cm$^{-3}$ \citep{mckee2018}. Refractive echoes have also been detected over months-long timescales \citep{backer2000,lyne2001}, and coincided with periods where the observed scattering deviated dramatically from the canonical scattering model \citep{backer2000,lyne2001}. Individual, giant pulses from the Crab also show evidence of multiple scattered trains \citep{sallmen99}. Changes in the scattering time could be correlated with orbital phase if the FRB source is in a binary system (which is one of the scenarios that could give rise to the large observed RM sign changes). Refraction through a companion outflow could also periodically enhance the burst flux \citep{johnston96,main2018}. All of these effects have been observed from Galactic pulsars \citep{johnston96,main2018,andersen2022} and may be observable from FRB 20190520B, although we have not detected them in the data set considered here. \\
\indent Regardless of the exact physical scenario, FRB local environments may not always yield burst structure consistent with the canonical scattering model typically assumed for burst shapes. Bursts' temporal structure can deviate from the canonical scattering model for several reasons: The exponential PBF applies to the special case of a Gaussian scattered image, but for non-Gaussian scattered images, the mean scattering delay will be greater than the $1/e$ time of an exponential PBF \citep{lambert99}. When the scattering screen is spatially well-confined (such as in a filament or discrete patch), the scattering strength is not uniform in directions transverse to the LOS, and the shape of the scattered image will be influenced by the physical extent of the screen rather than small ($\ll$ au) scale plasma density fluctuations \citep{cl_trunc_screen2001}. In this case, the frequency dependence of $\tau$ can be significantly shallower than $\nu^{-4}$, {and the scattering tail will be truncated} \citep{cl_trunc_screen2001}. We have identified two high S/N bursts from FRB 20190520B that {fall in Set 2} (do not show the frequency dependence expected from canonical scattering), but which have skewness functions with significant evidence of temporal asymmetries that may be related to scattering through a non-uniform screen {(these are the bursts shown in Figure~\ref{fig:maxskew3746} in Appendix~\ref{appendix:skew})}. These effects, combined with the degree of variability we have characterized using the canonical scattering model, suggest that scattering may be variable in other FRBs, including as yet one-off FRBs that may not be representative of their source's local scattering medium, and repeating FRBs that have not yet shown obvious scattering. Scattering variations may be detectable regardless of whether sources also show RM variations and PRSs. Future searches for scattering variations from other repeating FRBs, in addition to correlations between scattering, flux, DM, and polarization, will illuminate how sub-parsec scale processes in FRBs' local environments shape burst propagation and observed spectra. 

\section*{Acknowledgements}
The authors thank the anonymous referee, Paz Beniamini, and Jason Hessels for their comments on this work. SKO, JMC, and SC acknowledge support from the National Science Foundation (AAG-1815242) and are members of the NANOGrav Physics Frontiers Center, which is supported by NSF award PHY-2020265. CHN is supported by the FAST Fellowship and DL acknowledges support from the National Natural Science Foundation of China (NSFC) Programs No. 11988101 and No. 11725313. JWM gratefully acknowledges support by the Natural Sciences and Engineering
Research Council of Canada (NSERC), [funding reference \#CITA 490888-16]. CJL acknowledges support from the National Science Foundation under Grant No.~2022546. RAT acknowledges support from NSF grant AAG-1714897.

\section*{Data Availability} The FAST data used in this paper are available at \url{https://doi.org/10.11922/sciencedb.o00069.00004}.

\bibliographystyle{mnras}
\bibliography{bib}

\appendix

\section{Interpreting the Skewness Amplitude}\label{appendix:skew}

\begin{figure}
    \centering
    \includegraphics[width=0.45\textwidth]{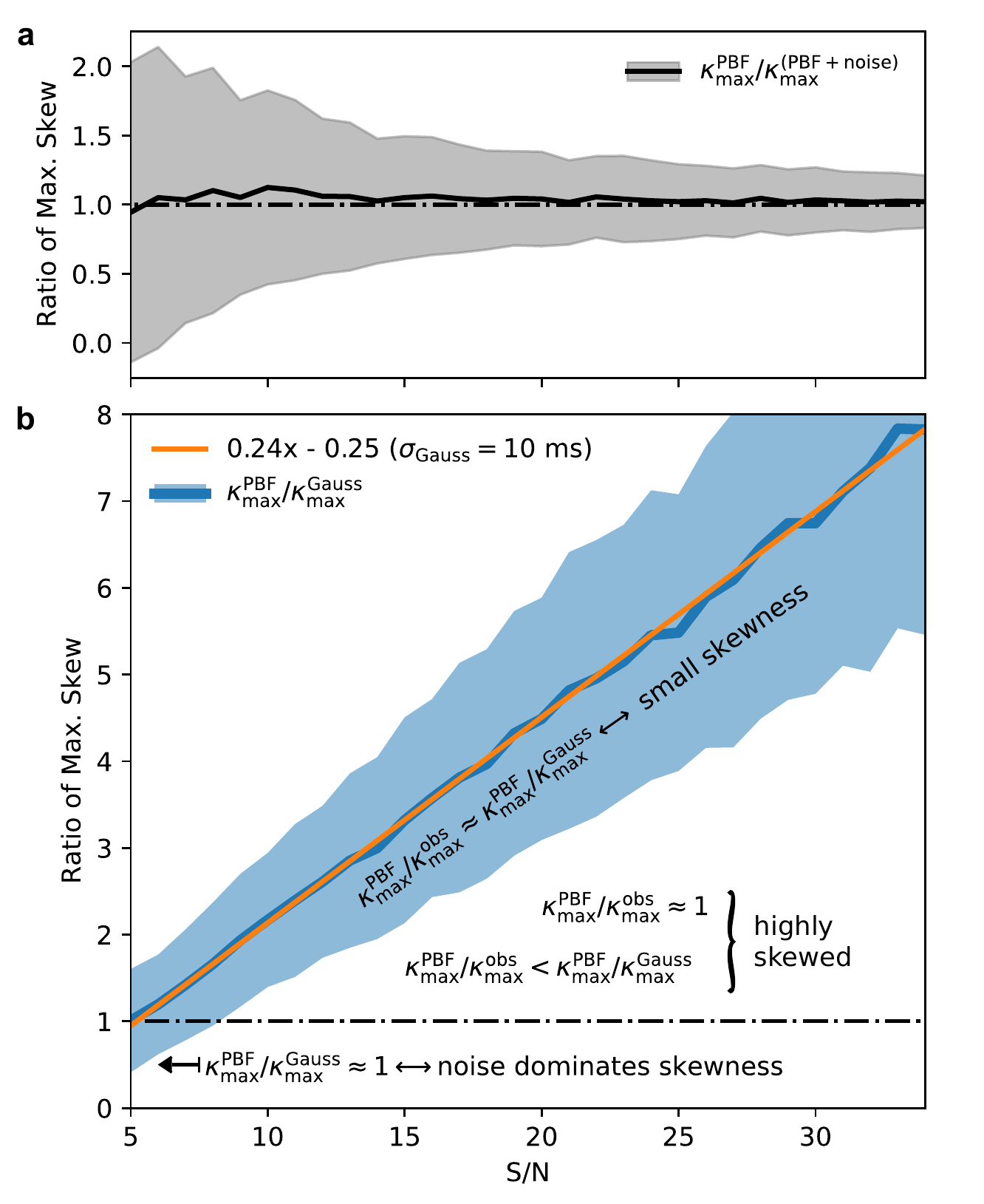}
    \caption{Simulated examples of the maximum skewness ratio for one-sided exponential and Gaussian pulses. a) Ratio of maximum skewness for a noiseless one-sided exponential PBF to a noisy PBF {($\kappa_{\rm max}^{\rm PBF}/\kappa_{\rm max}^{\rm (PBF+noise)}$)}, as a function of S/N. The solid black line and shaded grey region correspond to the mean and standard deviation of the maximum skewness ratio for 500 independent noise realizations. The dashed-dotted black line indicates where the ratio equals one. b) The solid blue line and shaded region show the mean and standard deviation, respectively, of the maximum skewness ratio for a noiseless exponential PBF to a noisy Gaussian with a standard deviation $\sigma_{\rm Gauss} = 10$ ms {($\kappa_{\rm max}^{\rm PBF}/\kappa_{\rm max}^{\rm Gauss}$)}. The solid orange line indicates the best-fit linear model, which scales with Gaussian standard deviation as $\sigma_{\rm Gauss}^{5.4}$. The dashed-dotted line indicates where the skewness ratio equals one. At high S/N values, {skewness ratios for observed bursts ($\kappa_{\rm max}^{\rm PBF}/\kappa_{\rm max}^{\rm obs}$) falling within the blue region correspond to bursts with very small skewness (no scattering)}. {High S/N bursts that have $\kappa_{\rm max}^{\rm PBF}/\kappa_{\rm max}^{\rm obs}\approx 1$ are highly skewed (evidence for scattering).} At low S/N values, noise dominates the skewness, and the maximum skewness of a noisy Gaussian pulse becomes indistinguishable from the maximum skewness of the PBF; {in this noise-dominated regime, $\kappa_{\rm max}^{\rm PBF}/\kappa_{\rm max}^{\rm obs}$ does not provide evidence of scattering}.}
    \label{fig:skewratio_sim}
\end{figure}

For noisy pulses with a range of unscattered and scattered widths, the amplitude of maximum skewness $\kappa_{\rm max}$ does not have a simple, deterministic relationship with scattering time $\tau$. We therefore assess whether $\kappa_{\rm max}$ for a given burst shows evidence of scattering by comparing the observed $\kappa_{\rm max}^{\rm obs}$ to the value $\kappa_{\rm max}$ would have if the burst were maximally asymmetric; i.e., if the entire burst width were contributed by the PBF. The ratio of maximum skew, $\kappa_{\rm max}^{\rm PBF}/\kappa_{\rm max}^{\rm obs}$, is then compared to the ratio that would be obtained for a Gaussian burst with the same observed total width and S/N, $\kappa_{\rm max}^{\rm PBF}/\kappa_{\rm max}^{\rm Gauss}$. This comparison of ratios is equivalent to testing whether an observed burst is distinguishable from a Gaussian, {to a given level of statistical confidence}. Figure~\ref{fig:skewratio_sim} shows that the maximum skew ratio $\kappa_{\rm max}^{\rm PBF}/\kappa_{\rm max}^{\rm Gauss}$ is a linear function of S/N. The mean and rms error in this skewness ratio is computed from 500 independent white noise realizations. At high S/N, $\kappa_{\rm max}^{\rm PBF} \gg \kappa_{\rm max}^{\rm Gauss}$, because the Gaussian pulse's skewness is small compared to the skewness of a PBF with the same total width. At low S/N, noise dominates the skewness function, and the ratio $\kappa_{\rm max}^{\rm PBF}/\kappa_{\rm max}^{\rm Gauss}$ approaches unity. In this regime, the skewness function of the noisy Gaussian is indistinguishable from the skewness of an equivalent-width PBF. The slope of $\kappa_{\rm max}^{\rm PBF}/\kappa_{\rm max}^{\rm Gauss}$ and the S/N at which it reaches unity depend on pulse width. Figure~\ref{fig:skewratio_sim} shows the best-fit linear model for $\kappa_{\rm max}^{\rm PBF}/\kappa_{\rm max}^{\rm Gauss}$ vs. S/N, for a Gaussian standard deviation $\sigma_{\rm Gauss} = 10$ ms. This linear model scales with Gaussian width as roughly $\sigma_{\rm Gauss}^{5.4}$, {based on simulations of the maximum skew ratio for Gaussian widths ranging between 5 ms and 40 ms}. 

\begin{figure}
    \centering
    \includegraphics[width=0.5\textwidth]{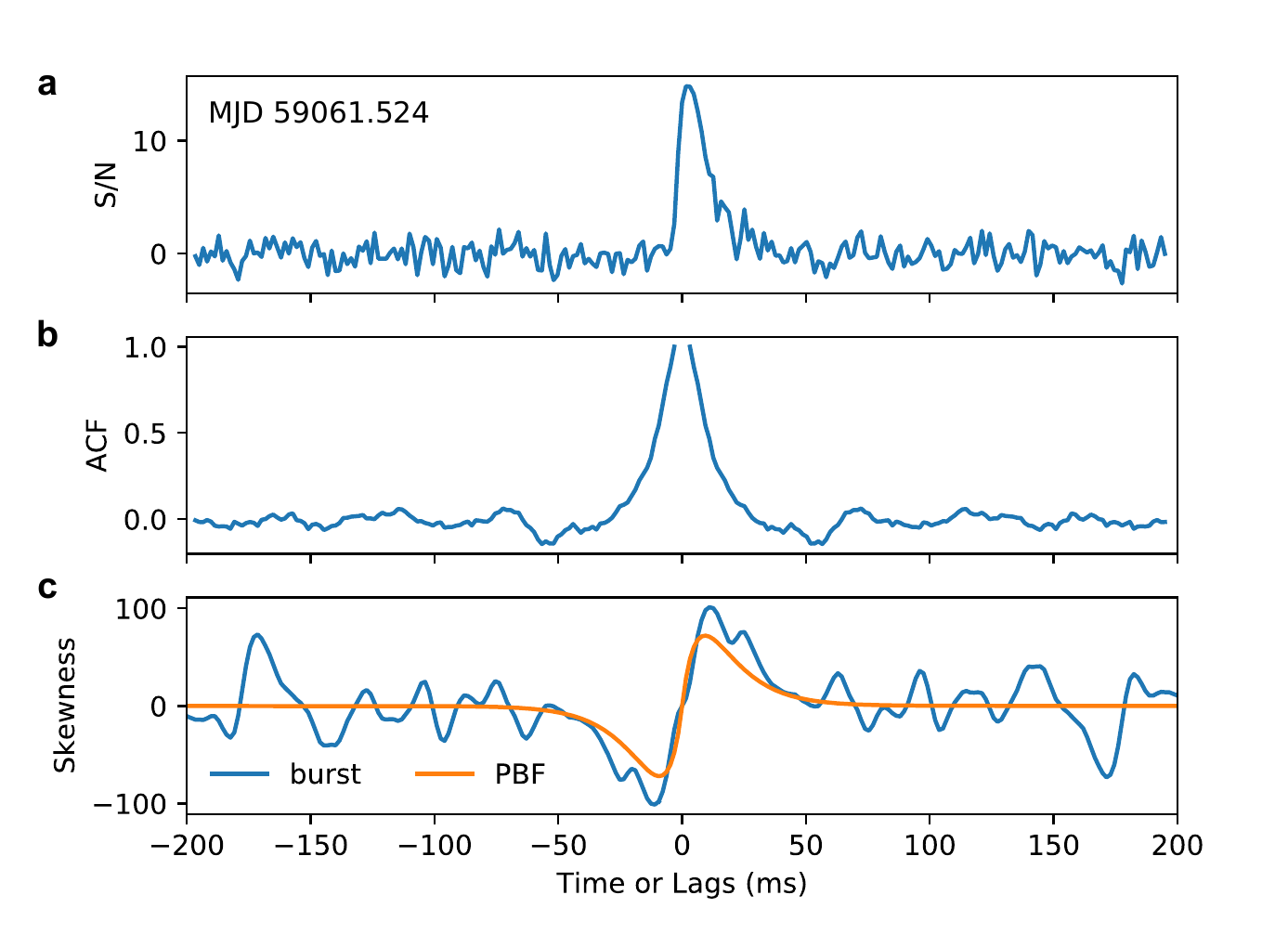}
    \caption{Intensity, autocorrelation, and skewness functions for a scattered burst. a) Frequency-averaged burst intensity vs. time in S/N units, for Burst A shown in Figure~\ref{fig:burst_set1}. This burst has a measured scattering time $\tau = 6.7\pm0.4$ ms at 1.45 GHz. b) Autocorrelation function (ACF) vs. time lag, calculated from the burst profile shown in panel (a). c) Skewness as a function of time lag for the measured burst profile (blue) and the skewness of a one-sided exponential pulse broadening function (PBF) with the same total width as the observed burst (orange). The total width was measured using the ACF. The ratio of the observed maximum skewness to the maximum skewness of the PBF is consistent with a very positively skewed burst {($>95\%$ confidence)}, as expected from scattering.}
    \label{fig:maxskew23}
\end{figure}

\begin{figure*}
    \centering
    \includegraphics[width=\textwidth]{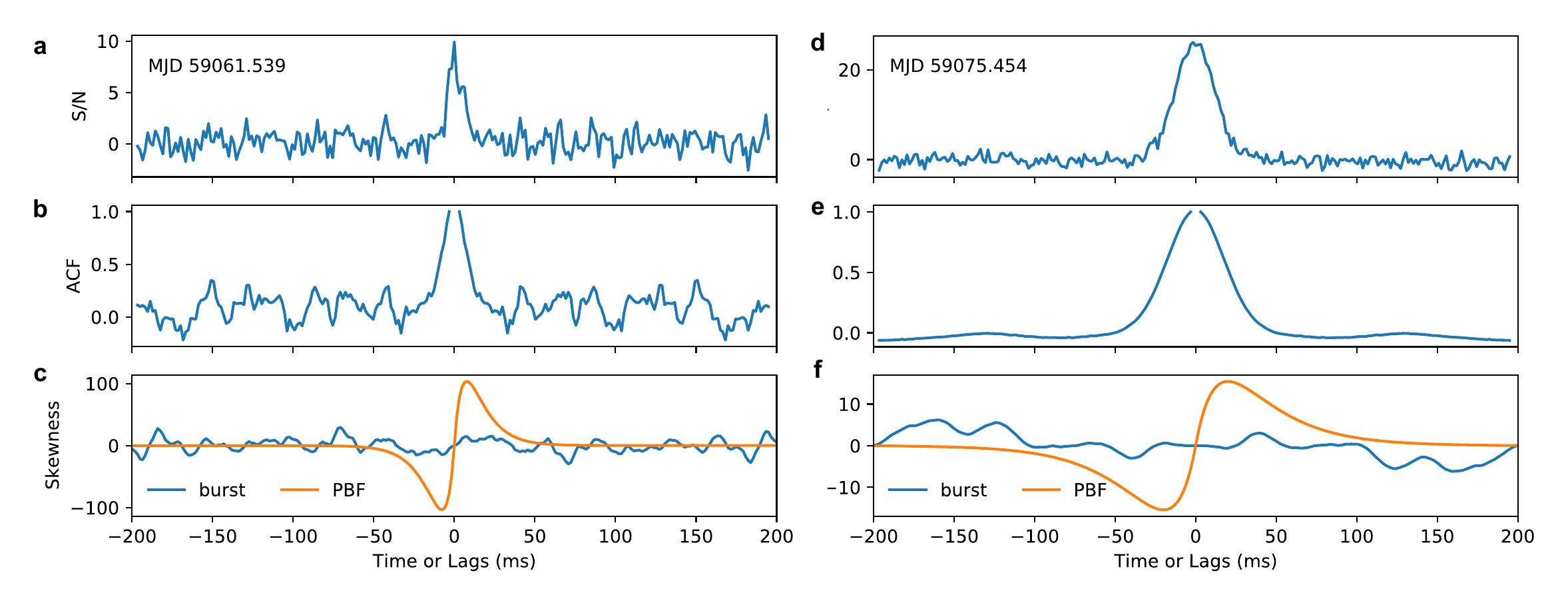}
    \caption{Intensity, autocorrelation, and skewness functions for two bursts without significant skewness. a) - c) Same as Figure~\ref{fig:maxskew23} for a burst detected at MJD 59061.539. The ratio of the observed maximum skewness to the maximum skewness of the PBF is consistent with skewness dominated by noise {($>95\%$ confidence)}, and the presence of scattering is indeterminate. d) - f) Same as a) - c) for a burst detected at MJD 59075.454. In this case, the S/N is high but the ratio of observed maximum skewness to the PBF maximum skewness is consistent with very small skewness {($>95\%$ confidence)}, and the presence of scattering is again indeterminate.}
    \label{fig:maxskew2845}
\end{figure*}

\begin{figure*}
    \centering
    \includegraphics[width=\textwidth]{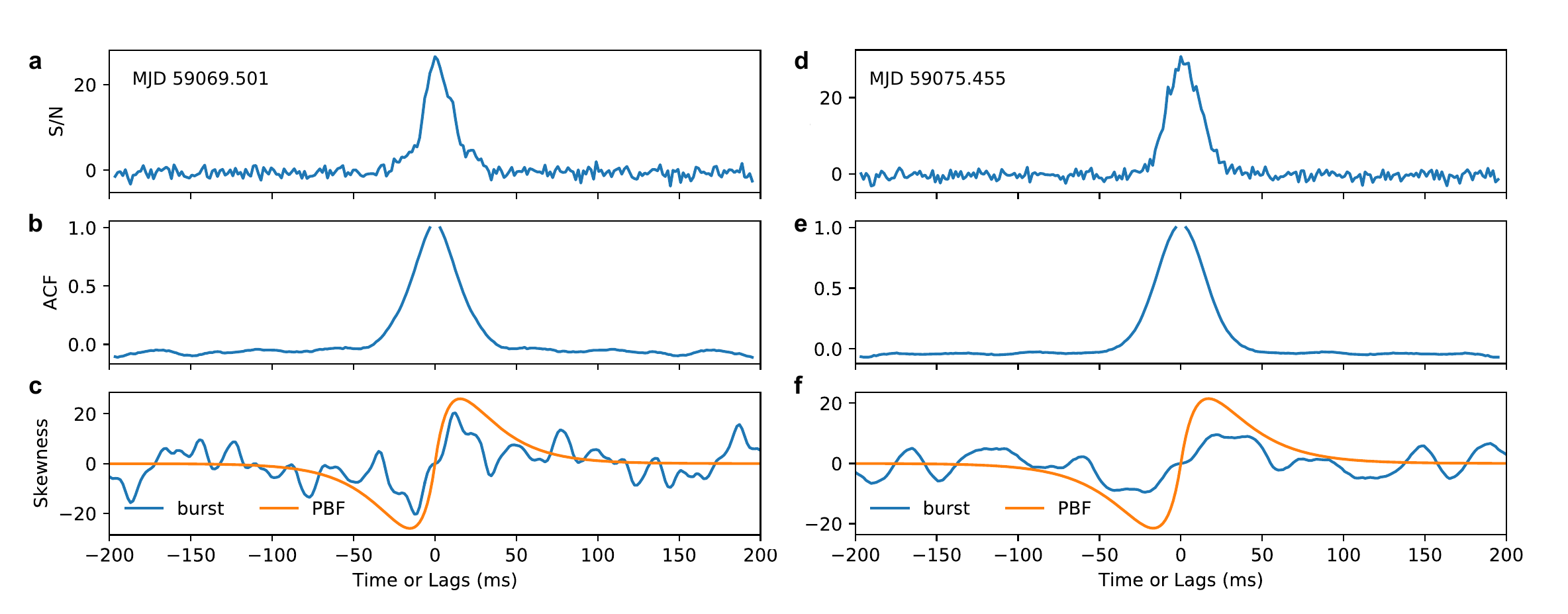}
    \caption{Intensity, autocorrelation, and skewness functions for two bursts with significant evidence of skewness. a) - c) Same as Figure~\ref{fig:maxskew23}a for a burst detected at MJD 59069.501. The ratio of the observed maximum skewness to the maximum skewness of the PBF is close to one, and is inconsistent with the ratio expected for a noisy Gaussian burst (see Figure~\ref{fig:skewratio_sim}) of the same width at $99\%$ confidence. d) - f) Same as a) - c) for a burst detected at MJD 59075.455. The ratio of the observed maximum skewness is again inconsistent with the ratio expected from a noisy Gaussian of the same width ($95\%$ confidence). The canonical scattering model did not yield significant constraints on $\tau$ for either of these bursts.}
    \label{fig:maxskew3746}
\end{figure*}

In practice, we calculate the ratio of maximum skewness for each observed burst $\kappa_{\rm max}^{\rm PBF}/\kappa_{\rm max}^{\rm obs}$, assuming an exponential PBF that has the same total width as measured from the burst ACF. This ratio is then compared to the simulated mean and rms error of $\kappa_{\rm max}^{\rm PBF}/\kappa_{\rm max}^{\rm Gauss}$ for the same S/N and total width. {We determine whether $\kappa_{\rm max}^{\rm PBF}/\kappa_{\rm max}^{\rm Gauss}$ falls into one of two relevant regimes:
\begin{enumerate}
    \item If $\kappa_{\rm max}^{\rm PBF}/\kappa_{\rm max}^{\rm Gauss} \leq 1$ to within $95\%$ confidence (based on the simulated error), the burst skewness falls in the noise-dominated regime, and the presence of scattering is considered indeterminate. 
    \item If, on the other hand, $\kappa_{\rm max}^{\rm PBF}/\kappa_{\rm max}^{\rm Gauss} > 1$ (to at least $95\%$ confidence), then the burst does not fall in the noise-dominated regime.
\end{enumerate}
In the second case, $\kappa_{\rm max}^{\rm PBF}/\kappa_{\rm max}^{\rm obs} \approx \kappa_{\rm max}^{\rm PBF}/\kappa_{\rm max}^{\rm Gauss}$ indicates that the burst skewness is smaller than expected from scattering, whereas $\kappa_{\rm max}^{\rm PBF}/\kappa_{\rm max}^{\rm obs} < \kappa_{\rm max}^{\rm PBF}/\kappa_{\rm max}^{\rm Gauss}$ indicates that the burst skewness is consistent with scattering, at a given confidence interval based on the simulated error in $\kappa_{\rm max}^{\rm PBF}/\kappa_{\rm max}^{\rm Gauss}$.} Figures~\ref{fig:maxskew23}-~\ref{fig:maxskew3746} show comparisons between the observed skewness and skewness of an equivalent-width PBF for a burst in Set 1 and four bursts in Set 2, which demonstrate cases where the observed skewness is both consistent and inconsistent with scattering. The two bursts shown in Figure~\ref{fig:maxskew3746} are cases where a fit for the canonical scattering model was indeterminate, but both $\delta t_{\rm max}$ and $\kappa_{\rm max}$ reveal significant temporal asymmetries { that may hint at scattering from a non-uniform screen ($\kappa_{\rm max}^{\rm obs}$ is highly skewed to $>99\%$ confidence for the lefthand burst and to $95\%$ confidence for the righthand burst in Figure~\ref{fig:maxskew3746})}.

\bsp	
\label{lastpage}
\end{document}